\documentclass[10pt,aps,twocolumn,superscriptaddress,floatfix,preprintnumbers,prd,nofootinbib]{revtex4-1}
	
\usepackage{array}[=2016-10-06]
    
\usepackage{subcaption}
\usepackage{ragged2e}
\usepackage{amsmath}
\DeclareCaptionJustification{justified}{\justifying}
\def\wzeta{\widehat{\zeta}}
\makeatletter
\newcommand{\subsetsim}{\mathrel{\mathpalette\subset@sim\relax}}
\newcommand{\subset@sim}[2]{%
  \vtop{\offinterlineskip\m@th
    \ialign{\hfil##\cr
      $#1\subset$\cr\noalign{\kern0.5pt}\scalebox{0.9}{$#1\sim$}\cr
    }%
  }%
}
\makeatother

\usepackage{multirow}
\usepackage{tabularx}
\usepackage{makecell}

\usepackage{amssymb,amsmath,verbatim,mathtools,needspace,enumitem,etoolbox,graphicx,physics,microtype,afterpage,bm}
\usepackage[dvipsnames, usenames]{xcolor}
\usepackage[normalem]{ulem}
\usepackage{booktabs}
\usepackage[unicode, colorlinks=true, linkcolor=linkcolor, citecolor=linkcolor, filecolor=linkcolor,urlcolor=linkcolor, pdfusetitle]{hyperref}
\usepackage[all]{hypcap}
\usepackage[T1]{fontenc}
\usepackage[utf8]{inputenc}
\usepackage{tabularx}
\usepackage{float}
\definecolor{darkpurple}{RGB}{48, 0, 72} 

\newcommand{\OGW}{\Omega_\text{\tiny GW}}
\newcommand{\rhoGW}{\rho_\text{\tiny GW}}
\newcommand{\vx}{{\bf x}}
\newcommand{\vk}{{\bf k}}

\newcommand{\calH}{\cal H}

\usepackage{multirow}
\usepackage{pifont}
\usepackage{lmodern}

\usepackage{orcidlink}

\allowdisplaybreaks
\usepackage{tikz}
\usepackage{color}
\usepackage{framed}
\usepackage{hyperref}
\hypersetup{colorlinks, citecolor=bluscuro, linkcolor=black, urlcolor=bluscuro}
\definecolor{rossos}{cmyk}{0,1,1,0.55}
\definecolor{bluscuro}{rgb}{0.15, 0.2, .85}
\definecolor{bluchiaro}{cmyk}{1,.3,0.,0.1}
\definecolor{ForestGreen}{rgb}{0.13, 0.55, 0.13}

\def\bea{\begin{eqnarray}}
\def\eea{\end{eqnarray}}

\newcommand{\bs}{\begin{subequations}}
\newcommand{\es}{\end{subequations}}

\newcommand{\be}{\begin{equation}}
\newcommand{\ee}{\end{equation}}
\renewcommand{\d}{{\rm d}}

\def\lsim{\mathrel{\rlap{\lower4pt\hbox{\hskip0.5pt$\sim$}}
    \raise1pt\hbox{$<$}}}         
\def\gsim{\mathrel{\rlap{\lower4pt\hbox{\hskip0.5pt$\sim$}}
    \raise1pt\hbox{$>$}}}         

\def\d{{\mathrm{d}}}

\newcommand{\cern}{Department of Theoretical Physics, CERN, Esplanade des Particules 1, P.O. Box 1211, Geneva 23, Switzerland}
\newcommand{\unipd}{Dipartimento di Fisica e Astronomia ``G. Galilei'', Università degli Studi di Padova, via Marzolo 8, I-35131 Padova, Italy}
\newcommand{\infnpd}{INFN, Sezione di Padova, via Marzolo 8, I-35131 Padova, Italy}
\newcommand{\unige}{D\'epartement de Physique Th\'eorique   and Gravitational Wave Science Center\\
Universit\'e de Gen\`eve, 24 quai E. Ansermet, CH-1211 Geneva, Switzerland}

\begin{document}

\preprint{CERN-TH-2026-015}

\title{
Dark Matter from Eternity
}

 \author{Gabriele Franciolini$^{\orcidlink{0000-0002-6892-9145}}$}
 \email{gabriele.franciolini@unipd.it}
 \affiliation{\unipd}
 \affiliation{\infnpd}
 \affiliation{\cern}

 \author{Marco Peloso$^{\orcidlink{0000-0002-9348-9970}}$}
 \email{marco.peloso@pd.infn.it}
 \affiliation{\unipd}
 \affiliation{\infnpd}

 \author{Antonio Riotto$^{\orcidlink{0000-0001-6948-0856}}$}
 \email{antonio.riotto@unige.ch}
\affiliation{\unige}

\date{\today}

\begin{abstract}
\noindent
We propose that the totality of dark matter in the universe might ascribe its origin to one of the key properties of cosmological inflation, that it may be eternal: regions that at the end of  the primordial accelerated expansion of the universe never reheated, but  keep  eternally inflating, manifest themselves as   primordial black holes   in our observable universe. This mechanism can provide a primordial black hole abundance which is larger than the standard one due to the gravitational collapse of sizeable overdensities in the radiation phase. 
It also predicts  a broad spectrum for the curvature perturbation  and  a flat  stochastic  gravitational wave background  at a level of $\Omega_\text{\tiny GW} h^2 \simeq 10^{-10}$ up to the  mHz. 
\end{abstract}

\maketitle


\noindent{{\bf{\em Introduction.}}} 
\label{sec:introduction}
The  composition of Dark Matter (DM) in our universe constitutes one of the fundamental and open questions in
physics \cite{Arcadi2023ParticleDM}. One  possibility is that  DM is composed by Primordial Black Holes (PBHs) \cite{Carr:2026hot}. This option is exciting because it would not necessarily invoke any physics beyond the Standard Model with the only exception of a period of accelerated expansion, inflation,  during the early stages of the evolution of the universe \cite{Lyth:1998xn}. PBHs might indeed be generated when sizeable  curvature perturbations, originated during inflation, re-enter the horizon and collapse. This process is also accompanied by the production of a stochastic background of Gravitational Waves (GWs) sourced by the curvature perturbations themselves at second-order in perturbation theory \cite{Tomita:1975kj,Matarrese:1992rp} (for a review, see Ref. \cite{Domenech:2021ztg}).

In this letter we present a novel mechanism to produce the PBHs which might comprise the totality of the DM in our universe. It can provide a PBH abundance larger than the one generated by   the standard mechanism, which  relies on the collapse of large overdensities during the radiation phase. In fact, the PBHs we will deal with due their origin to an intrinsic property of cosmological inflation, the fact that some patches might   eternally inflate \cite{Linde:1986fc}. We will therefore dubb such PBHs as DM  from eternity. 

During inflation, the   quantum fluctuations of the curvature of spacetime generates  randomness:   different regions, with comoving   size of  the order of the  Hubble radius during inflation, behave differently; 
some spots  stop inflating and reheat, 
other spots with sizeable values of the curvature perturbation keep inflating forever. If at the end of inflation in our observable universe, there are patches which are still eternally inflating, they are perceived by an external observer as  BHs since
the eternally inflating region 
remains  hidden behind the BH Schwarzschild horizon \cite{Blau:1986cw}. Even though these regions are rare, the implications are profound as the totality of the DM might indeed ascribe its origin to such eternally inflating patches if the resulting  PBHs have an asteroid-like mass.   

DM from eternity requires large fluctuations of the curvature perturbation and therefore is  inevitably accompanied by a large amount of stochastic GWs and we will show that its spectrum must be  broad. This is an observational bonus: the scenario will be tested in the forthcoming space-based experiment LISA in the mHz range \cite{LISACosmologyWorkingGroup:2025vdz}, and possibly as well in experiments probing smaller frequencies.

\vskip 0.2cm
\noindent{{\bf{\em  Regions of eternal inflation in our observed universe.}}} 
\label{sec:BH form}
During inflation curvature perturbations are generated with an amplitude which freezes as soon as the corresponding physical wavelength encompasses the comoving Hubble radius $H^{-1}$.  On the CMB scales the power spectrum ${\cal P}_\zeta (k)$ of such fluctuations  is flat and of the order of $10^{-9}$.  Imagine, however, that  the power spectrum of the curvature perturbation is enhanced for some period toward the end of inflation by some mechanism whose origin  we remain agnostic about,  with ${\cal P}^{1/2}_\zeta\simeq 10^{-1}$ or so. Our observable universe is composed of roughly $e^{3N}$ independent regions of size $H^{-1}$ generated during inflation, where  $N\simeq 60$  corresponds to our observable universe. The central question is whether out of this humongous number of regions, in some of them inflation never stops and becomes eternal. The criterion for this to happens is that the curvature perturbation acquires a random value larger than unity, 
\be
\zeta\gsim \zeta_\text{\tiny c}= 1.
\ee
Indeed, a region inflates eternally if the quantum fluctuations of the clock driving inflation becomes more sizable than the classical displacement. Calling $\phi$ such  a clock (the inflaton), quantum fluctuations are 
\be
\Delta \phi_\text{\tiny q}=\frac{H}{2\pi},
\ee
while the classical displacement in a Hubble time is 
\be
\Delta \phi_\text{\tiny cl}=\frac{\dot\phi}{H}.
\ee
Imposing $\Delta \phi_\text{\tiny q}\gsim \Delta \phi_\text{\tiny cl}$ gives
\be
\frac{H^2}{2\pi\dot\phi}\equiv \zeta\gsim 1= \zeta_\text{\tiny c},
\ee
where in the first equality we have identified the curvature perturbation in the flat gauge. 

The  fraction of  regions of comoving size $\sim H^{-1}$ contained in our observable universe at the end of inflation which end up eternally inflating is given by  
\be \label{a}
\beta_\text{\tiny eter}=\int_{\zeta_\text{\tiny c}}^\infty\frac{{\rm d} \zeta }{\sqrt{2\pi}\sigma_\zeta^2}e^{-\zeta^2/2\sigma_\zeta^2}\approx\frac{\sigma_\zeta}{\sqrt{2\pi}\zeta_\text{\tiny c}} e^{-\zeta_\text{\tiny c}^2/2\sigma_\zeta^2},
\ee
where the approximation holds for the physically relevant case of $\sigma_\zeta \ll 1$ and we have used a Gaussian probability for the curvature perturbation.  This can be a good approximation. For instance, in  ultra-slow-roll models the curvature perturbation $\zeta$ is connected to its Gaussian component $\zeta_{\text{\tiny g}}$ by a relation of the form \cite{Cai:2018dkf,Atal:2019cdz,Biagetti:2021eep,Caravano:2025diq}

\be
\zeta=-\mu\,{\rm ln}\left|1-\frac{\zeta_{\text{\tiny g}}}{\mu}\right|,
\ee
where the non-Gaussianity is parametrized by $f_{\text{\tiny NL}}=(5/6\mu)$ and  $\zeta_{\text{\tiny g}}$ is the Gaussian component. For $\mu\gg  1$, one indeed has $\zeta\simeq \zeta_{\text{\tiny g}}$ and  $f_{\text{\tiny NL}}\ll 1$.

The variance depends on the shape of the enhanced power spectrum of the curvature perturbation. In the case of a sharply peaked power spectrum, ${\cal P}_\zeta (k)\simeq A_{\text{\tiny p}} k_*\delta(k-k_*)$, one obtains $\sigma_\zeta^2=A_{\text{\tiny p}}$; while for the case of a broad power spectrum, ${\cal P}_\zeta (k)  \simeq A_{\text{\tiny b}} \theta(k_\text{\tiny max}-k) \theta(k-k_\text{\tiny min})$, one obtains 
\be
\sigma_\zeta^2=A_{\text{\tiny b}}\ln\left(\frac{k_\text{\tiny max}}{k_\text{\tiny min}}\right).
\ee

\vskip 0.2cm
\noindent{{\bf{\em Primordial black holes from eternity.}}} 
\label{sec:eternity}
 Suppose now that regions of size of the comoving Hubble radius where eternal inflation is occuring do exist at the end of inflation in our observable patch. Such regions will never reheat. However, an observer (that is,  us) in the exterior true-vacuum region (that is approximately Minkowski today) will describe that region  as a BH, while an observer in the interior will describe a closed universe which completely disconnects from the original spacetime \cite{Blau:1986cw,Garriga:2015fdk}.  In other words, the eternally inflating region 
remains  hidden behind the BH Schwarzschild horizon because of  the
distortion of the metric.
 This means that today our observable universe might be populated by PBHs whose mass is dictated by the energy contained in a comoving Hubble volume during inflation.  If, for simplicity,   such regions are mainly generated right before the end of inflation (see for instance Ref. \cite{Leach:2000yw}), the PBH mass reads
 \be
 \label{aaa}
M_\text{\tiny PBH}\simeq \frac{M^2_\text{\tiny Pl}}{H}\simeq 10^{-1}\frac{M_\text{\tiny Pl}^3}{T^2_\text{\tiny RH}}\simeq 10^{-12}\left(\frac{10^5\,{\rm GeV}}{T_\text{\tiny RH}}\right)^2 M_\odot,
 \ee
where $M_\text{\tiny Pl}$ is the reduced Planckian mass, $M_\odot$ is the solar mass, $T_\text{\tiny RH}$ is the reheating temperature and we have assumed instantaneous reheating at the end of inflation for simplicity with a total number of relativistic degrees of freedom of about $10^2$. 
The present abundance of dark matter in the form of PBHs per logarithmic mass interval  is given by
\be
\label{fPBH}
f_\text{\tiny PBH}(M_\text{\tiny PBH})\simeq \left(\frac{\beta_\text{\tiny eter}}{6\cdot 10^{-9}}\right)\left(\frac{M_\odot}{M_\text{\tiny PBH}}\right)^{1/2},
\ee
which can be order unity for $\sigma_\zeta^2\simeq 0.017$ and $M_\text{\tiny PBH}\simeq 10^{-12} M_\odot$. This value is in a narrow mass window for which PBHs can constitute the totality of the dark matter of the universe, without conflicting with observational bounds \cite{Carr:2026hot}.

This mechanism of PBH from eternity is well distinct from the standard mechanism of formation from the collapse of overdensities when the perturbations re-enter the horizon. Both mechanisms are at play in our scenario. In the following we study their relative efficiency and determine when the PBH abundance is dominated by those formed during inflation. 

\vskip0.2cm
\noindent{{\bf{\em PBH abundance from the standard  collapse in a radiation phase.}}} 
\label{sec:radiation}
Regions of the Universe that experience large curvature perturbations but do not enter the eternal inflation regime eventually re-enter the Hubble horizon after reheating. Upon horizon re-entry during radiation domination, these perturbations can collapse into PBHs through the standard mechanism driven by large radiation overdensities \cite{Suyama2025}. In this case, the relevant quantity controlling PBH formation is the compaction function \cite{Shibata:1999zs}, which captures the integrated mass excess within a given radius and is constructed from spatial derivatives of the primordial curvature perturbation.

At super-Hubble scales and assuming spherical symmetry, the density contrast can be expressed in terms of the curvature perturbation as~\cite{Harada:2015yda}
\begin{equation}
\label{eq:SphericalDelta}
\delta(r,t) = 
-\frac{4}{9}
\left(\frac{1}{aH}\right)^2 
e^{-2\zeta(r)}\left[
\zeta^{\prime\prime}(r) + \frac{2}{r}\zeta^{\prime}(r) + \frac{1}{2}\zeta^{\prime 2}(r)
\right],
\end{equation}
where $^\prime \equiv \mathrm{d}/\mathrm{d}r$. Substituting this expression into the definition of the compaction function (see e.g. Ref. \cite{Musco:2020jjb}) leads to%
\begin{equation}
\label{eq:CompactionFull}
\mathcal{C}(r) = 
-\frac{4}{3}\,r\,\zeta^{\prime}(r)\left[
1 + \frac{r}{2}\zeta^{\prime}(r)
\right]
\equiv
\mathcal{C}_1(r) - \frac{3}{8}\mathcal{C}_1^2(r).
\end{equation}
Since the density contrast depends on spatial derivatives of $\zeta$, the variance of the compaction function is insensitive to a constant curvature plateau. Assuming the absence of strong non-Gaussianities correlating long and short scales, its variance is given by
\begin{equation}
\label{eq:Var1}
\sigma_{\text{\tiny c}}^2 =
\frac{16}{81}\int_0^{\infty}\frac{\mathrm{d}k}{k}
(kr_m)^4
T^2(k,r_m)
W^2(k,r_m)\,
{\cal P}_\zeta(k),
\end{equation}
where 
\be
T(k,\eta ) \hspace{-0.1cm}=\hspace{-0.1cm} \frac{3}{(k \eta /\sqrt{3})^3}\left[\sin(k \eta /\sqrt{3})\hspace{-0.1cm}-\hspace{-0.1cm}(k \eta /\sqrt{3})\cos(k \eta /\sqrt{3})\right]\hspace{-0.1cm}, 
\label{transfer}
\ee
and
$W(k,r)=3[\sin(kr)-kr\cos(kr)]/(kr)^3$ 
(notice that the transfer and  window functions are absent in the evaluation of the PBHs from eternal inflation because the production is happening during inflation and on a volume of size $\sim H^{-1}$).

Under these circumstances, the probability distribution of the compaction function reduces to a Gaussian,
\begin{equation}
\label{eq:PDFCompa2}
\mathcal{P}_{\text{\tiny g}}(\mathcal{C}_1) =
\frac{1}{\sqrt{2\pi}\,\sigma_{\text{\tiny c}}}
\exp\!\left(
-\frac{\mathcal{C}_1^2}{2\sigma_{\text{\tiny c}}^2}
\right),
\end{equation}
which directly controls the abundance of PBHs formed at Hubble re-entry. One finds
\be
\label{aa}
    \beta_{\text{\tiny rad}} = \int_{\mathcal{C}_\text{\tiny c}}^\infty\d \mathcal{C}P(\mathcal{C})=
    \int_{\mathcal{C}_{1\text{\tiny c}}}^\infty\d \mathcal{C}_1P_{\text{\tiny g}}(\mathcal{C}_1)\simeq \frac{\sigma_{\text{\tiny c}}}{\sqrt{2\pi}\mathcal{C}_{1\text{\tiny c}}} e^{-\mathcal{C}^2_{1\text{\tiny c}}/2\sigma_{\text{\tiny c}}^2},
 \ee   
where 
\be
\mathcal{C}_{1\text{\tiny c}}=\frac{4}{3}\left(1-\sqrt{\frac{2-3\mathcal{C}_\text{\tiny c}}{2}}\right).
\ee
As a reference value  we take $\mathcal{C}_\text{\tiny c}\simeq 0.55$ \cite{Musco:2020jjb}, for which $\mathcal{C}_{1\text{\tiny c}}\simeq 0.77$.

For a sharply peaked power spectrum one has $k_* r_m\simeq 2.7$ \cite{Germani:2018jgr}, and $\sigma_{\text{\tiny c}}^2\simeq 1.2\sigma_\zeta$. Since
$\mathcal{C}_{1\text{\tiny c}}<\zeta_{\text{\tiny c}}$ and $\sigma_{\text{\tiny c}}^2>\sigma_\zeta^2$, the abundance of PBH from eternity is always smaller than the one from the collapse during the radiation phase.

However, for a flat spectrum, one has $k_\text{\tiny max} r_m\simeq 3.5$ \cite{Germani:2018jgr} and 
\be
\sigma_{\text{\tiny c}}^2\simeq 0.98\, A_{\text{\tiny b }}=0.98\ln^{-1}\left(\frac{k_\text{\tiny max}}{k_\text{\tiny min}}\right)\sigma_\zeta^2.
\ee
Comparing the exponentials in Eqs. (\ref{a}) and (\ref{aa}),
\be
\frac{\mathcal{C}^2_{1\text{\tiny c}}}{2\sigma_{\text{\tiny c}}^2}
=\frac{(0.77)^2}{2\cdot 0.98\,\sigma_\zeta^2}\ln\left(\frac{k_\text{\tiny max}}{k_\text{\tiny min}}\right)=0.6\ln\left(\frac{k_\text{\tiny max}}{k_\text{\tiny min}}\right)\frac{\zeta^2_\text{\tiny c}}{2\sigma_\zeta^2},
\ee
shows that the abundance of the PBHs from the collapse during radiation is smaller than the one from eternity,
\be
\beta_\text{\tiny rad}\lsim \beta_\text{\tiny eter}\,\,\,\,\,\, {\rm for}\,\,\, \,\,\, \ln\left(\frac{k_\text{\tiny max}}{k_\text{\tiny min}}\right)\gsim 1.6,
\ee
a very mild condition for the duration of the phase in which the broad power spectrum is generated (it corresponds to more than  about   one e-fold and a half) right before  the end of inflation. In such a case, since the variance grows with the spread between $k_\text{\tiny min}$ and $k_\text{\tiny max}$, the largest abundance of PBHs corresponds to the wavenumber $k_\text{\tiny max}$ and therefore to the mass in Eq.~(\ref{aaa}) if $k_\text{\tiny max}\simeq H$, as previously assumed.

\vskip 0.2cm
\noindent{{\bf{\em Induced gravitational waves.}}} 
\label{sec:induced} 
Due to the nonlinear nature of gravity, the enhanced scalar perturbations responsible for PBHs also inevitably source GWs~\cite{Tomita:1975kj,Matarrese:1992rp,Acquaviva:2002ud,Mollerach:2003nq,Ananda:2006af,Baumann:2007zm}. This production is accurately captured at second-order in perturbation theory, where a GW mode is generated by the coupling of two enhanced scalar modes obtained from the linear theory—namely, those we have considered so far. In the following, we present the starting definitions, a brief description of the computation, and the result for the case of our interest. Details of the computation can be found for instance in~\cite{Baumann:2007zm}. We start from the line element 
\begin{equation}
\mathrm d s^2 = -a^2  
\left( 1 + 2 \Psi \right) \mathrm d\eta^2 + a^2\left[ \left( 1 - 2 \Psi \right) \delta_{ij} + \frac{1}{2}h_{ij}\right] \mathrm dx^i \mathrm dx^j ,
\label{line}
\end{equation}
where $\delta_{ij}$ is the Kronecker delta, and where the GW is encoded by the $h_{ij}$ tensor fluctuations, which is constrained to be transverse and traceless: $\partial_i h_{ij} = h_{ii} = 0$. Finally, $\Psi$ denotes the scalar perturbation (in the Newtonian-gauge and in absence of significant anisotropic stress), which is related to the curvature perturbation $\zeta \left( \vk \right)$ imprinted on super-horizon scales during inflation by
\begin{equation}
\Psi \left( \eta ,\, \vk \right) = \frac{2}{3} T \left( k \eta \right) \zeta \left( \vk \right) \,, 
\end{equation}
with the transfer function $T$ acquiring the form~\eqref{transfer} in the radiation dominated era. Einstein equations in the geometry~\eqref{line} result in the evolution equation for the tensor mode
\begin{equation}
h_{ij}'' + 2\mathcal H h_{ij}' - \nabla^2 h_{ij} = -4 \mathcal T_{ij}{}^{lm} \mathcal S_{lm},
\label{eq: eom GW1}
\end{equation}
where prime denotes differentiation with respect to conformal time, ${\cal H} \equiv a'/a$ is the conformal Hubble rate, and where $\mathcal T_{ij}{}^{lm}$ projects out the transverse-traceless component of the source. During the radiation dominated epoch, the source, evaluated to second-order in the scalar mode, reads~\cite{Acquaviva:2002ud} 
\begin{equation}
\mathcal S_{lm} = 2 \partial_l \partial_m \Psi^2 - 2 \partial_l \Psi \partial_m \Psi - \partial_l \left( \frac{\Psi'}{\calH} + \Psi \right) \partial_m \left( \frac{\Psi'}{\calH} + \Psi \right) \,.
\end{equation} 
We decompose the tensor perturbation $h_{ij}$ over the linear polarization basis $\{e^{(+)}_{ij}, e^{(\times)}_{ij}\}$ (see e.g.~ Ref. \cite{Baumann:2007zm} for details)
\begin{equation}
h_{ij}(\eta, \vx)\hspace{-0.1cm} =\hspace{-0.1cm} \int \hspace{-0.1cm}\frac{ d^3k}{(2\pi)^3} \left[h_{\vk}^{(+)}(\eta)e_{ij}^{(+)}(\vk) \hspace{-0.1cm}+ \hspace{-0.1cm}h_{\vk}^{(\times)}(\eta)e_{ij}^{(\times)}(\vk)\right] e^{i\vk \cdot \vx} \,, 
\end{equation}
solve Eq. ~\eqref{eq: eom GW1} in Fourier space, and evaluate the energy density $\rhoGW = (M^2_\text{\tiny Pl}/16 \pi a^2) \left\langle h_{ij}' h_{ij}' \right\rangle$. The energy density is conventionally expressed in terms of the fractional energy density per logaritmic interval
\begin{equation}
\Omega_\text{\tiny GW} (k) \equiv \frac{1}{\rho_\text{\tiny cr}} \, \frac{\d \rhoGW}{\d \ln k} \;, 
\end{equation}
where $\rho_\text{cr}$ is the critical energy of a flat universe (as observations indicate that our universe is extremely flat~\cite{Planck:2018vyg}, the integral of $\OGW$ in $\d \ln k$ essentially gives the ratio between the GW energy density and the total energy density of the universe). 

The scalar-induced gravitational wave spectrum emerging from a number of scalar profiles can be found for instance in~\cite{LISACosmologyWorkingGroup:2025vdz}, which presents the prospect of detecting such a signal in LISA. The GW energy density is quadratic in the gravitational wave, and therefore is proportional to the expectation value of the four point function of the curvature perturbation $\zeta$. Assuming Gaussian linear scalar perturbations in the source, $\OGW$ will then be proportional to two powers of the scalar power spectrum, namely to $A_{\text{\tiny b}}^2$ in our context. For a flat curvature power spectrum, one finds 
\cite{Kohri:2018awv}
\begin{align}
\Omega_\text{\tiny GW} h^2 
& \simeq 
0.82\, A_{\text{\tiny b}} ^2 \,
\Omega_\text{\tiny r,0} h^2 
 \simeq 
1.6 \times 10^{-9} \left(\frac{A_{\text{\tiny b}}}{10^{-2}}\right)^2, 
\end{align}
where $\Omega_\text{\tiny r,0} h^2  = 4.2 \times 10^{-5}$ is the present energy density in relativistic species (conventionally normalized to include also the contribution of neutrinos, as if they were massless today).

\begin{figure}[t]
    \centering
    \includegraphics[width=\linewidth]{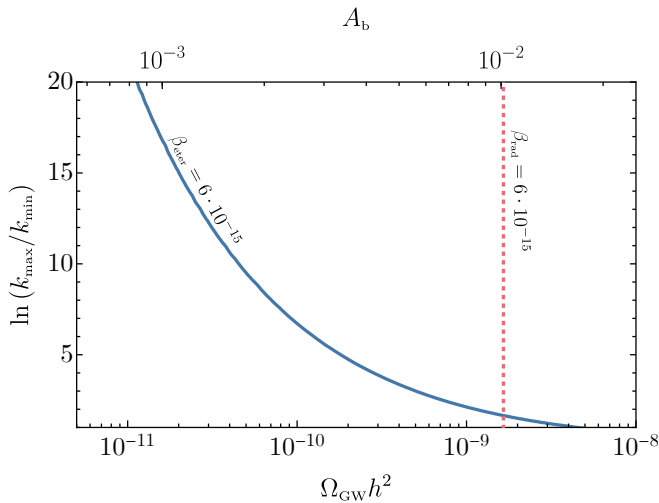}
    \caption{GW abundance associated with PBH DM with asteroidal masses, for both the new mechanism discussed in this letter and the one assuming collapse in a radiation-dominated universe. On the vertical axis we vary the width of the variance of the curvature perturbation. }
    \label{plotOmega}
\end{figure}

From Eq. (\ref{fPBH}) we have seen that the totality of the DM might be composed by PBHs from eternity if $\sigma^2_\zeta=A_{\text{\tiny b}}\ln(k_\text{\tiny max}/k_\text{\tiny min})\simeq 0.017$.  If the duration of the broad spectrum is of order unity in terms of e-folds, we see that the prediction in terms of the stochastic GW background may fall inside the LISA sensitivity band with an amplitude $\Omega_\text{\tiny GW} h^2\simeq 10^{-10}$. We show this in Fig.~\ref{plotOmega}.
The corresponding stochastic GW spectrum has    maximum frequency  
\be
f_\text{\tiny GW,max}\simeq \left(\frac{ 10^{-12}\,M_\odot}{M_\text{\tiny PBH}}\right)^{1/2}\,{\rm mHz}.
\ee
 Furthermore, the minimum frequency 
is given by $f_\text{\tiny GW,min}=(k_\text{\tiny min}/k_\text{\tiny max})f_\text{\tiny GW,max}$. The  broadness of the GW spectrum becomes  therefore a phenomenological probe of the duration of the inflationary phase during which the curvature perturbation is enhanced with respect to the CMB value. We notice, however, that going down to the nHz frequency would require $(k_\text{\tiny max}/k_\text{\tiny min})$=(mHz/nHz)$=10^6$, or $\ln(k_\text{\tiny max}/k_\text{\tiny min})\simeq 14$. This gives $A_{\text{\tiny b}}\simeq (\sigma_\zeta^2/14)\simeq 10^{-3}$, if we insist in the totality of DM from eternity. In such a case, 
$
\Omega_\text{\tiny GW} h^2 
 \simeq  
 10^{-11}$, 
still visible in the mHz range \cite{LISACosmologyWorkingGroup:2025vdz}, but not by the pulsar timing array  experiments in the nHz range (see, for instance, Ref. \cite{Cecchini:2025oks}).

\vskip 0.2cm
\noindent{{\bf{\em Conclusions.}}} 
\label{sec:conclusions}
In this Letter we have proposed that DM might due its origin to one of the fundamental  properties  of cosmological inflation, the fact that regions of spacetime might still eternally inflating today. Being hidden behind the Schwarzschild horizon, in our observed universe they are detected  as BHs. We have shown that their abundance can comprise the totality of the DM if the PBH mass is asteroid-like.  This possibility comes with an observational gift.  If DM is made of PBHs from eternally inflating regions, an inescapable prediction of such a scenario is the fact that a stochastic GW background will be detected in the mHz range, possibly extending to significantly lower frequencies.

There are many avenues of improvement. First of all, one might extend our reasoning to the case in which the curvature perturbation is not Gaussian at very large values (see, for instance, Ref. \cite{Escriva:2023uko}), possibly making use of the extreme value statistics along the line of Ref. \cite{Franciolini:2026jat}. Secondly, other PBH masses might be considered, less interesting for the DM abundance, but more relevant for their role in BH mergers and their detectability in GW ground-based experiments. Finally, one might relax the assumption of instant reheating, by introducing some level of model dependence.

\vskip1cm
\centerline{\bf Acknowledgments}
\vskip 0.3cm
\noindent
We thank V. De Luca and A.J. Iovino for interesting discussions. M.P. and G.F. acknowledge support from Istituto
Nazionale di Fisica Nucleare (INFN) through the Theoretical Astroparticle Physics (TAsP)
project. M.P. additionally acknowledges support from the European Union’s Horizon 2020 research
and innovation programme under the Marie Sklodowska-Curie grant agreement
No. 101086085-ASYMMETRY. A.R.  acknowledges support from the  Swiss National Science Foundation (project number CRSII5\_213497).

\appendix


\twocolumngrid
\bibliographystyle{JHEP}
\bibliography{draft}

\end{document}